# A computational framework for integrating Predictive processes with evidence Accumulation Models (PAM)

Antonino Visalli[1*], Francesco Maria Calistroni[2], Margherita Calderan[1], Francesco Donnarumma[3], Marco Zorzi[1,4], Ettore Ambrosini[1,2]

[1] Department of General Psychology and Padova Neuroscience Center, University of Padova, Padova (IT)

[2] Department of Neuroscience, University of Padova, Padova (IT)

[3] Institute of Cognitive Sciences and Technologies, National Research Council, Rome (IT)

[4] IRCCS San Camillo Hospital, Venice (IT)

**\*Corresponding author:**

Antonino Visalli

Department of General Psychology, University of Padova

Via Venezia 8 - 35131 Padova (IT)

antonino.visalli@unipd.it

**ABSTRACT**

Evidence Accumulation Models (EAMs) have been widely used to investigate speeded decision-making processes, but they have largely neglected the role of predictive processes emphasized by theories of the predictive brain. In this paper, we present the Predictive evidence Accumulation Models (PAM), a novel computational framework that integrates predictive processes into EAMs. Grounded in the "observing the observer" framework, PAM combines models of Bayesian perceptual inference, such as the Hierarchical Gaussian Filter, with three established EAMs (the Diffusion Decision Model, Lognormal Race Model, and Race Diffusion Model) to model decision-making under uncertainty. We validate PAM through parameter recovery simulations, demonstrating its accuracy and computational efficiency across various decision-making scenarios. Additionally, we provide a step-by-step tutorial using real data to illustrate PAM's application and discuss its theoretical implications. PAM represents a significant advancement in the computational modeling of decision-making, bridging the gap between predictive brain theories and EAMs, and offers a promising tool for future empirical research.

# INTRODUCTION

Evidence accumulation models (EAMs), also known as sequential sampling models, are a class of cognitive models for behavioral responses in speeded decision-making tasks. A prominent example is the Diffusion Decision Model (DDM; Forstmann et al., 2016), in which a dichotomous response is generated through a process of noisy evidence accumulation (i.e., diffusion) beginning at a starting point and continuing until a decision threshold associated with a specific choice is reached (Figure 1). Other EAM examples include models such as the Lognormal Race Model (LNR; Rouder et al., 2015), Linear Ballistic Accumulator Model (Brown & Heathcote, 2008), and Race Diffusion Model (RDM; Tillman et al., 2020), which assume separate racing accumulators for each possible choice. Irrespective of their different assumptions and specificities, EAMs present two main advantages. First, in modelling responses, they account for both response time (RT) and response choice jointly. Second, EAMs are defined by various parameter types that can be meaningfully connected to latent cognitive or neural processes (Lewandowsky & Farrell, 2011; Voss et al., 2004). As an illustration, consider the four key parameters of the basic DDM (Ratcliff, 1978): the drift rate ($v$), which represents the average slope of the evidence accumulation process; the boundary separation ($a$), which is the distance between the two thresholds; the starting point ($z$), which denotes the location between the two thresholds from which the accumulation process starts; and the non-decision time ($Ter$), which determines a temporal shift in the diffusion process. These parameters have compelling theoretical interpretations with empirical support (Ratcliff & McKoon, 2008; Voss et al., 2004). The absolute value of $v$ is usually interpreted in terms of speed in information processing (Schmiedek et al., 2007) and is often associated with task difficulty or variability in individual cognitive abilities (Voss et al., 2004). On the other hand, $a$ is considered a measure of the amount of information required to make a decision and has been mapped to task strategies (Boywitt & Rummel, 2012), such as the speed-accuracy trade-off (Voss et al., 2004). Concerning *z*, it reflects an *a priori* bias towards one decision over the other, which has been observed for asymmetries in the likelihood that one decision is correct or in its reward (Mulder et al., 2012; Voss et al., 2004). Finally, *Ter* encompasses the time needed for non-decisional processes, including stimulus encoding and, especially, motor processes (Gomez et al., 2015; Lerche & Voss, 2016; Voss et al., 2004).

As the labels "evidence accumulation models" and "sequential sampling models" suggest, most of the research employing this computational approach has described decision-making by focusing on stimulus-driven (bottom-up) processes. However, this is not sufficient for theoretical frameworks considering the brain as a predictive machine (Clark, 2013). Leading theories of information processing, such as predictive coding or the Bayesian brain hypothesis (Doya et al., 2007; Friston, 2010), indeed, assume that the brain does not passively process sensory information but instead has internal generative models of the environment, which are used to infer the hidden causes of events. The brain thus actively predicts incoming signals and probabilistically encodes information in terms of posterior probabilities derived by integrating "likelihoods" of newly gained information with "prior" beliefs. Although the idea of a predictive brain has become highly influential in cognitive (neuro)science, it has had little impact on the EAM field. With some exceptions (Bitzer et al., 2014; Fontanesi et al., 2019; Frank, 2013; Fu et al., 2022; Pedersen et al., 2017; Summerfield & de Lange, 2014), indeed, the role of predictive processes has been essentially neglected. This limited integration between predictive brain theories and EAMs poses a significant constraint to our understanding of decision-making mechanisms. Bridging this gap could yield more comprehensive models that account for both bottom-up and top-down influences on decision-making.

To foster research on predictive processes in speeded decision making, we introduce PAM (Predictive evidence Accumulation Models), a new computational framework that integrates predictive processes in EAMs. PAM offers a flexible, efficient and valid analytical tool, which is accessible even to researchers with limited experience in computational modeling.

The development of PAM has been grounded on the "observing the observer" framework (Daunizeau, den Ouden, Pessiglione, Kiebel, Friston, et al., 2010; Daunizeau, den Ouden, Pessiglione, Kiebel, Stephan, et al., 2010), a generic modeling approach of inferential mechanisms and decision-making. This framework (Figure 1) assumes that an agent is provided with a trial-wise sequence of task inputs to which it responds by producing a sequence of behavioral outputs. The agent attempts to infer the hidden environmental causes of its sensory inputs. Inference rests on a *perceptual* model with some learning parameters. The agent's behavior depends on inferred beliefs and is described by a *decision* or *response* model. As stated by Daunizeau and colleagues (2010), the central idea of this approach is to make inferential (predictive) processes part of decision processes. The "observing the observer" is a meta-inferential framework since it enables experimenters to make statistical inference about participants' perceptual inference. Indeed, once perceptual and decision models are defined, experimenters can use task inputs and behavior to estimate both the perceptual and response parameters of the participants.

In the present article, we implement and validate three PAMs that implement the "observing the observer" framework by combining a specific decision model - here we consider the DDR, LNR, and RDM - with a perceptual model based on the Hierarchical

Gaussian Filter (HGF; Mathys et al., 2011, 2014). The HGF, it is a generic computational model of approximate (variational) Bayesian learning under (perceptual) uncertainty and environmental volatility that has been widely used for the empirical investigation of predictive processes in different cognitive domains, including perception (Stefanics et al., 2018; Visalli, Capizzi, et al., 2023), associative learning (Iglesias et al., 2013; Marshall et al., 2016), attention (Vossel et al., 2014), cognitive control (Visalli, Ambrosini, et al., 2023; Viviani et al., 2024), and social cognition (Diaconescu et al., 2014). Nevertheless, in the Github repository ([github.com/antovis86/PAM-PredictiveAccumulationModels](github.com/antovis86/PAM-PredictiveAccumulationModels)), we also provide alternative PAMs implementing a different perceptual model (i.e., the Volatile Kalman Filter; Piray & Daw, 2020).

In what follows, we first describe the three implementations of PAMs. Secondly, we present a validation of PAMs using parameter recovery via simulations. Finally, we provide a step-by-step tutorial on using PAM with a real dataset.

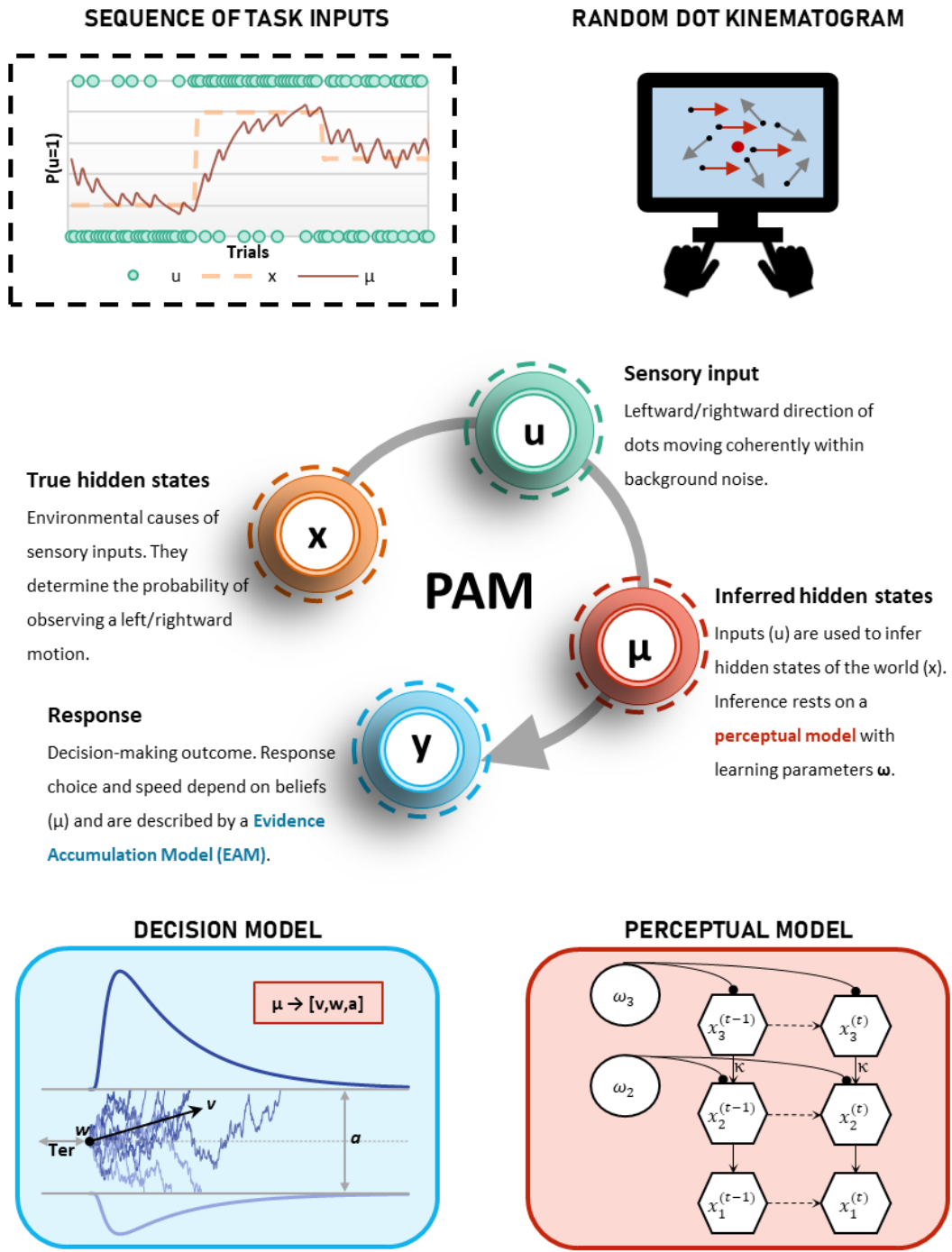

**Figure 1. Infographic overview of PAM.** PAM is grounded in the "Observing the Observer" framework, which posits that agents attempt to infer hidden environmental causes of their sensory inputs. In this example, participants perform a Random Dot Kinematogram task, where they are required to judge the leftward or rightward direction of dots moving coherently within background noise. To optimize behavior, participants infer the probability of observing rightward motion. This inference relies on a perceptual model with individual learning parameters. The perceptual model depicted in the figure mirrors the Hierarchical Gaussian Filter. The inferred states obtained through the inversion of the perceptual model guide decision-making, which is described

by an Evidence Accumulation Model. In the proposed example, a Drift Diffusion Model is used as the decision model. Its parameters (the drift rate "v", the starting point "w", and boundary separation "a") are trial-wise modulated by inferred states from the perceptual model.

## MODEL SPECIFICATIONS

As mentioned in the introduction, the development of PAM has been grounded in "observing the observer" framework (Daunizeau, den Ouden, Pessiglione, Kiebel, Stephan, et al., 2010), which is based on the specification of two distinct components: a p*erceptual model* and a *decision model*. The perceptual model embodies the set of beliefs that an agent holds about the environment, functioning as a mapping from sensory inputs to the agent's representations of hidden causes. On the other hand, the decision model acts as a mapping from the hidden causes to the behavioral response. Specifically, here, we illustrate our framework using the HGF for binary inputs as the perceptual model (Mathys et al., 2011, 2014) in combination with three EAMs we have redesigned as decision models: the DDM (Ratcliff, 1978), the LNR (Rouder et al., 2015) and the RDM (Tillman et al., 2020). The code for the HGF is part of the Translational Algorithms for Psychiatry-Advancing Science (TAPAS) software package (Frässle et al., 2021). All the models have been implemented using the MATLAB programming language.

In the next sections, we first provide a description of the perceptual model and the decision models used, along with the corresponding modeling choices. Particular attention is given to how the decision models are influenced by the prior beliefs obtained from the perceptual model. Finally, an overview of the optimization procedures and fitting methods will be presented.

**Perceptual model**

*Hierarchical Gaussian Filter*

The HGF is a perceptual model that tracks a participant's beliefs about the task structure across multiple hierarchical levels (see "Perceptual Model" in Figure 1). Specifically, we used the HGF for binary inputs (Mathys et al., 2011). At the lowest level, the model encodes the binary state $x_1^{(t)}$, whose value represents whether at trial $t$ the target $u$ takes the value of 1 or 0. Since $x_1$ is a binary variable, its probability is determined by a single real variable - the state $x_2$ encoded at the middle level - according to a logistic sigmoid (softmax) function. It follows that when $x_2 = 0$, the states $x_1 = 1$ and $x_1 = 0$ are equally probable; for $x_2 \rightarrow \infty$, the probability for $x_1 = 1$ and $x_1 = 0$ approaches 1 and 0, respectively; conversely, for $x_2 \rightarrow -\infty$, the probability for $x_1 = 1$ and $x_1 = 0$ approaches 0 and 1, respectively. The HGF model assumes that the value of $x_2$ changes across trials as a Gaussian random walk. The value $x_2^{(t)}$ is normally distributed around its value at the preceding trial ($x_2^{(t-1)}$), and with variance $exp(\kappa x_3^{(t)} + \omega_2)$. This implies that changes in $x_2$ over trials are determined by a state $x_3$ encoded at the highest level, as well as by a fixed (participant's specific) parameter $\omega_2$ which can capture individual differences in belief updating about $x_2$. In other terms, the state $x_3^{(t)}$, which represents the environmental log-volatility at trial $t$, and the stable participant-specific $\omega_2$ determine together the speed of learning at the middle level (how much the beliefs about $x_2$ can change from trial to trial). The parameter $\kappa$ determines the strength of the coupling between the second and the third levels. As for $x_2$, $x_3^{(t)}$ performs a Gaussian random walk, that is, its value is normally distributed around $x_3^{(t-1)}$ and with variance $exp(\omega_3)$. Here, $\omega_3$ is a fixed agent-specific parameter that represents meta-volatility (log-volatility of volatility). Trial-wise updating of posterior densities (i.e., beliefs) about hidden states (i.e., $x_2$ and $x_3$) is achieved through variational

(approximate) Bayesian inversion under a mean field approximation of the perceptual model. For a detailed description, please refer to the original papers (Mathys et al., 2011, 2014).

**Decision models**

*Drift Diffusion Model*

To build a DDM for the "observing the observer", we capitalized on the Wiener First Time Passage (WFPT) model (Gondan et al., 2014; Navarro & Fuss, 2009). WFPT describes the evidence accumulation process as a Wiener diffusion process, a type of stochastic process denoted as *X(T)*. At time $T_0$, the process starts at *X(0) = z*. The dynamic of the process is described by a stochastic differential equation (Gondan et al., 2014; Navarro & Fuss, 2009):

$$\frac{d}{dt}X(T) \sim Normal(v, \sigma^2) \quad (1)$$

Where *v* represents the drift rate and $\sigma^2 = 1$ represents the variance. The stochastic process moves between two absorbing barriers, typically with the lower barrier set at 0 and the upper barrier set at *a*, with *0 < z < a*. The Wiener Diffusion model is thus parametrized by the drift parameter *v*, the boundary separation *a* and the start point *z*. For convenience, the relative starting point is defined as *w = z/a*, such that *0 < w < 1*. The probability that the process reaches the lower boundary at time T is expressed by the following formula:

$$f(v, a, w) = \frac{1}{a^2}exp\left(-vaw - \frac{v^2 T}{2}\right) f(0, 1, w) \quad (2)$$

It is straightforward to obtain the probability density function for the upper boundary by setting *v′ = -v* and *w′ = 1 - w* (Gondan et al., 2014). When integrated over *T*, the two probabilities yield the choice probability of the corresponding boundary. The probability density of the lower and upper boundary sum to *1*. Therefore, when combined, the two densities jointly represent the probability of both RT and decision choice. In the implementation of our model, we used the computationally efficient approximation of the WFPT provided by Navarro and Fuss (2009).

To integrate the evolving beliefs from the HGF model into our decision model, we modulate the three parameters of the WFPT, *w*, *a* and *v,* based on the trial-by-trial beliefs. This modulation is implemented in a linear regression-like fashion, where the effect of prior beliefs on the parameters of the decision model is captured through a combination of an intercept term and a coefficient for the predictions. While the exact implementation varies across parameters, our approach generally aims to capture the influence of these beliefs on the decision-making process. This approach ensures that the influence of the predicted beliefs about $x_1$ (denoted as $\hat{\mu} \in (0,1)$) on the EAM parameters can be estimated bilaterally, allowing for both positive and negative modulations relative to a baseline value. This property enables possible subsequent statistical analyses (see the Tutorial section).

Starting with the *w* parameter, its intercept is fixed to .5. Predicted beliefs are centered around .5, and their influence on trial-wise *w* are determined by the slope parameter $b_w$:

$$w^{(t)} = 0.5 + b_w \cdot \left(\hat{\mu}^{(t)} - 0.5\right) \quad (3)$$

Notably, if $b_w$ is 0, prior beliefs have no effect on the starting point, if $b_w$ is greater than 0, *w* moves towards the most probable barrier. If $b_w$ is less than 0, we obtain a reversed effect of probability, with *w* moving towards the less probable barrier. To constrain *w* between the two barriers, $b_w$ can assume values in the range (-1, 1).

For the *a* parameter, its intercept value $a_a$ is modulated by the precision (inverse of the variance $\hat{\mu} \cdot (1-\hat{\mu})$) of the predicted belief, scaled by a $b_a$ parameter:

$$a^{(t)} = a_a + b_a \cdot \left((s(\hat{\pi}^{(t)} - 4) - .5\right) \quad (4)$$

Specifically, we used the sigmoid transformation of precision $\hat{\pi}$ proposed in Vossel and colleagues (2014). Moreover, we subtracted .5 to set the intercept of *a* at the lowest level of precision (i.e., $\hat{\mu} = .5$). Similarly to $b_w$, if $b_a$ is 0, precision in the predictions has no effect on the boundary separation; while for negative values of $b_a$, the higher is the precision, the smaller the distance between the boundaries (*vice versa* for positive $b_a$ values).

Lastly, concerning the *v* parameter, the sign of its intercept value $a_v$ is determined by *u*. Similar to the starting point, prior beliefs are centered around .5 and their influence on *v* is determined by the slope parameter $b_v$ as follows:

$$v^{(t)} = I(u^{(t)} = 1) \cdot \left(a_v + b_v(\hat{\mu}^{(t)} - .5)\right) -$$

$$-I(u^{(t)} = 0) \cdot \left(a_v + b_v((1 - \hat{\mu}^{(t)}) - .5)\right) \quad (5)$$

In Eq. 5, $I(u^{(t)})$ represents the indicator function for the stimulus identity at trial *t*. Note that when $u = 0$, the complementary of $\hat{\mu}$ is used to modulate the drift. Again, if $b_v$ is 0, the probability has no effect on the drift, while if $b_v$ is higher than 0, the drift increases with increased probability of $u$ at trial t and vice-versa when $b_v$ is less than 0.

In addition to the five parameters already described, the model includes a sixth parameter *Ter*, which is subtracted from RT to obtain the decision time (T). This parameter is constrained in the range (0, min(RT)).

From Equations 3,4,5, we obtain trial-wise triplet of WFPT parameters that can be used to compute the probability of observed T for the taken decisions:

$$P(T^{(t)}) = WFPT(T^{(t)}|v^{(t)}, w^{(t)}, a^{(t)}, u^{(t)}) \quad (6)$$

*Lognormal Race Model*

The LNR model provides an alternative framework for decision-making compared to the DDM model (Rouder et al., 2015). Unlike the DDM, the LNR describes the process of decision-making as a race between two (or more) accumulators, each representing a potentially different choice. Evidence is accumulated linearly until one of the accumulators reaches the decision threshold, and then a response is given. The finishing time *y* of each accumulator follows a lognormal distribution, naturally bounded on the positive axis and characterized by its right-skewness. Formally, the distribution of y of the i-th accumulator at the trial *t* is defined as (Rouder et al., 2015):

$$y_i^{(t)} \sim Lognormal(\theta_i, \sigma_i^2) \quad (7)$$

Consequently, at trial *t*, the accumulator with the smallest finishing time determines the choice and T.

It follows that the joint probability *g* of choice *c* equal to *m* at time T can be formally defined as:

$$g(c = m, T) = f(T; \theta_m, \sigma_m^2) \prod_{i \neq m} \left(1 - F(T; \theta_i, \sigma_i^2)\right) \quad (8)$$

where *f* denotes the lognormal probability density function and *F* represents the lognormal cumulative distribution function. The joint density expresses the likelihood that the accumulator *m* reached the threshold at time T multiplied by the likelihoods that all the other accumulators ($i \neq m$) did not reach the threshold at time T.

For the LNR, the influence of the perceptual model's beliefs pertains only the mean of the lognormal distribution, $\theta$:

$$\theta_{c1}^{(t)} = a + b_{val}\left(I(u^{(t)} = 1)\right) + b(\hat{\mu}^{(t)} - .5) \quad (9)$$

$$\theta_{c0}^{(t)} = a + b_{val}\left(I(u^{(t)} = 0)\right) + b\left((1 - \hat{\mu}^{(t)}) - .5\right) \quad (10)$$

In Equations 9 and 10, $\theta_{c1}$ and $\theta_{c0}$ represent the means of the lognormal distributions for the first and second accumulators, which are associated with the choices corresponding to the binary stimulus values (*c1* for u = 1) and (*c0* for u = 0), respectively. The parameter *a* represents the intercept value of the means of the lognormal distributions for both accumulators. The slope $b_{val}$ adds to the intercept based on the accumulator validity (the valid accumulator is associated with the choice matching the stimulus value). Prior beliefs of *u* equal to the accumulator choice (i.e., $\hat{\mu}$ and 1-$\hat{\mu}$ for the first and second accumulator, respectively) are centered around .5. This centering ensures that deviations from .5 have proportional effects on both accumulators, but in opposite directions. The magnitude of this effect is modulated by the slope parameter *b*. If *b* is 0, there is no influence of prior beliefs on the mean of the lognormal distribution. If *b* is smaller than 0, the higher is the probability of *u* equal to the accumulator choice, the lower is the accumulator mean (and *vice versa*). Conversely, if *b* is greater than 0, the prior beliefs of *u* equal to the accumulator choice proportionally increase the accumulator mean.

To obtain the jointly probability of observed choice and T at trial *t*, the means of the two accumulators are assigned to *f* and *F* of Equation 8 as follows:

$$\theta_f^{(t)} = I(c^{(t)} = 1)\theta_{c1}^{(t)} + I(c^{(t)} = 0)\theta_{c0}^{(t)} \quad (11)$$

$$\theta_F^{(t)} = I(c^{(t)} = 1)\theta_{c0}^{(t)} + I(c^{(t)} = 0)\theta_{c1}^{(t)} \quad (12)$$

In the formula above, $c^{(t)}$ represents the subject's choice on trial *t* (1 or 0) and $I(c^{(t)})$ represents the indicator function. If the choice is 1, the mean of the winning accumulator $\theta_f$ corresponds to $\theta_{c1}$ and the mean of the losing accumulator $\theta_F$ corresponds to $\theta_{c0}$, and *vice versa*.

*Racing Diffusion Model*

The third model we adapted for PAM is the RDM (Tillman et al., 2020). The RDM combines elements from both the DDM model and the LNR. Similar to the DDM, the RDM incorporates distinct parameters into decision-making processes. Conversely, like the LNR, the RDM involves a race among multiple accumulators, each representing distinct response alternatives. Each accumulator's finishing time distribution follows a Wald distribution (Wald, 1947), parametrized by *a* (decision boundary) and *v* (drift rate). Mathematically, the first passage of time distribution can be written as:

$$f(a, v) = a\sqrt{\frac{1}{2\pi T^3}} exp\left[-\frac{1}{2T}(vT - b)^2\right] \quad (13)$$

As for the LNR (Equation 8), the probability that one accumulator wins the race can be computed as follows:

$$g(c = m, T) = f(T; a_m, v) \prod_{i \neq m}(1 - F(T; a_m, v)) \quad (14)$$

Like the LNR and the DDM, our Racing Diffusion Model incorporates the perceptual model's beliefs to modulate specific parameters. In this model, both the decision threshold $a$ and the drift rates $v$ are influenced by the prior beliefs.

Starting with the $a$ parameter, the integration of prior beliefs is formulated as follows:

$$a_{c1}^{(t)} = a_a + b_a(\hat{\mu}^{(t)} - .5) \quad (15)$$

$$a_{c0}^{(t)} = a_a + b_a\left((1 - \hat{\mu}^{(t)}) - .5\right) \quad (16)$$

Here, $a_{c1}$ and $a_{c2}$ represent the decision thresholds for the two accumulators. The $a_a$ parameter represents the intercept for both thresholds, while $b_a$ modulates the influence of prior beliefs on the threshold. Prior beliefs of $u$ equal to the accumulator choice (i.e., $\hat{\mu}$ and 1-$\hat{\mu}$ for the first and second accumulator, respectively) are centered around .5. This centering ensures that deviations from .5 have proportional effects on both accumulators, but in opposite directions. If $b_a$ is 0, there is no influence of prior beliefs on the decision thresholds. If $b_a$ is smaller than 0, the higher is the probability of $u$ equal to the accumulator choice, the lower is the decision threshold (and *vice versa*). Conversely, if $b_a$ is greater than 0, the prior beliefs of $u$ equal to the accumulator choice proportionally increase the decision threshold of the accumulator.

Notably, the modulation of *a* in the RDM has a similar effect to shifting the starting point *w* in the DDM. In the RDM, when *a* is modulated based on prior beliefs, this adjustment effectively lowers the threshold for one accumulator, reducing the required evidence, while raising it for the other accumulator, increasing the required evidence. Similarly, in the DDM, shifting *w* reduces the amount of evidence required for one boundary, while increasing the amount of evidence required for the other boundary. Moreover, it is important to note that modulating the boundary separation *a* in the DDM cannot reflect prior beliefs in the same way as the modulation of *a* in the RDM. This is because the DDM boundary separation symmetrically affects both decision alternatives, increasing or decreasing the evidence required for either response.

The drift rates for the two accumulators are calculated as follows:

$$v_{c1}^{(t)} = a_v + b_{val}\left(I(u^{(t)} = 1)\right) + b_v(\hat{\mu}^{(t)} - .5) \ (17)$$

$$v_{c0}^{(t)} = a_v + b_{val}\left(I(u^{(t)} = 0)\right) + b_v((1 - \hat{\mu}^{(t)}) - .5) \ (18)$$

In Equations 17 and 18, $v_{c1}$ and $v_{c2}$ represent the drift rates for the first and second accumulator. The parameter $a_v$ represents the intercept value of drift rates. The slope $b_{val}$ adds to the intercept based on the accumulator validity (the valid accumulator is associated with the choice matching the stimulus value). The slope $b_v$ adjusts the drift rates based on the prior beliefs $\hat{\mu}$. In Equations 15-18, prior beliefs of *u* equal to the accumulator choice (i.e., $\hat{\mu}$ and 1-$\hat{\mu}$ for the first and second accumulator, respectively) are centered around .5. This centering ensures that deviations from the .5 baseline have effects on both accumulators, but in opposite directions.

Similar to the LNR, to obtain the joint probability of observed choice and T at trial *t*, the thresholds and drifts of the two accumulators are assigned to *f* and *F* of Equation 14 as follows:

$$a_f^{(t)} = I(c^{(t)} = 1)a_{c1}^{(t)} + I(c^{(t)} = 0)a_{c0}^{(t)} \quad (19)$$

$$a_F^{(t)} = I(c^{(t)} = 1)a_{c0}^{(t)} + I(c^{(t)} = 0)a_{c1}^{(t)} \quad (20)$$

$$v_f^{(t)} = I(c^{(t)} = 1)v_{c1}^{(t)} + I(c^{(t)} = 0)v_{c0}^{(t)} \quad (21)$$

$$v_F^{(t)} = I(c^{(t)} = 1)v_{c0}^{(t)} + I(c^{(t)} = 0)v_{c1}^{(t)} \quad (22)$$

**Optimization and Parameter Estimation**

Our approach to model fitting and parameter estimation was consistent across all three decision models, allowing for systematic comparison and robust integration with the perceptual component.

The model fitting procedure is based on the one implemented in the HGF toolbox (Frässle et al., 2021). The optimization algorithm used is the BFGS quasi-Newton optimization algorithm. This algorithm performs a joint optimization of both the perceptual model and the decision model. Its objective function finds the maximum-a-posteriori (MAP) estimates of the parameters for both the perceptual model and the decision model. Specifically, for the decision models, the procedure attempts to maximize the sum of the joint log-likelihood of observed RT and choice across all trials.

In the fitting procedure, optimization is performed using an unconstrained quasi-Newton algorithm, allowing it to search for parameter values across the entire real number line, i.e., (-

∞,+∞). This unconstrained search space is employed to simplify the optimization process, as certain optimization algorithms (like the BFGS method) perform more efficiently when there are no hard constraints. However, many model parameters have natural or theoretical limits (e.g., parameters that must be positive, or those that must lie within a specific range). To ensure the optimized parameters respect these inherent constraints, once the optimal values are found, they are transformed back to their respective native spaces using specific mapping functions. This step is critical for two reasons: it ensures that the parameter values are meaningful within the context of the model, and it maintains the integrity of the model's theoretical assumptions. For all models, the non-decision time parameter (*Ter*) is mapped back to its native space using a sigmoid function, bounded between 0 and the minimum observed RT. In the DDM, the intercepts $a_a$ and $a_v$ are mapped to the range (0, +∞) after optimization using an exponential function, ensuring positive starting point and valid drift rates. The slope $b_w$ is constrained within the range (-1, +1). In the LNR, the parameters remain unconstrained, meaning that no additional transformations are needed beyond the search space. In the RDM, the intercepts $a_a$ and $a_v$, are restricted to (0, +∞).

It is important to note that the PAM framework offers flexibility in model fitting. Users can decide for joint model fitting, which integrates both the perceptual and decision models, allowing for a comprehensive estimation of the entire cognitive process. Alternatively, the framework also supports the independent fitting of the decision model alone, using standard log-likelihood estimation. This flexibility is particularly useful when the goal is to focus solely on decision-making processes without explicitly modeling the perceptual stage.

**PARAMETER RECOVERY**

It is good practice to conduct a parameter recovery study when designing a new computational model in cognitive science, to assess the consistency of parameter estimation procedures (Heathcote et al., 2015; Wilson & Collins, 2019). The core idea of parameter recovery analysis is to generate synthetic data based on a set of parameters and then fit a model to the generated data to verify if the estimated parameters match the original ones. To validate our joint model, we conducted a parameter recovery analysis. A parameter recovery simulation was performed for each different decision model. For all the parameter recovery analyses, we used the same general methodologies, although some specific details changed, according to the models and the different sets of parameters. Finally, it is advisable to conduct a parameter recovery study for any future extension of the PAM model that the reader might implement.

**Data simulation**

*Simulated beliefs*

The trial-wise sequence of task inputs, used to simulate behavioral datasets, was the same as that employed in the real task experiment described in the step-by-step tutorial section. It consisted of twelve blocks in which the proportion of $u = 1$ was either .2, .5, or .8. The block sequence was designed to have triplets of blocks, ensuring an equal number of transitions from .2-blocks to .8-blocks and vice versa. Furthermore, in the real task, a 30-second pause was provided every three blocks. To avoid a bias towards 1 or 0 after the break, which might decay during the pause, each triplet ended with the .5-block. The input sequence within each block was obtained using the software MIX (van Casteren & Davis, 2006) with the aim of keeping the proportion of inputs stable throughout the block. Specifically, both the .2 and .8-blocks consisted of sequences of 35 inputs, where the minimum distance between two rare events (i.e., $u = 1$ for the .2-block and $u = 0$ for the .8-block) was 4. The .5-blocks consisted of sequences of 30 inputs that satisfied four constraints: any of the four possible sequences of two inputs could occur a maximum of 8 times; a maximum of three identical inputs in a row was allowed; any possible sequence of three inputs could occur three times, with the exception of the sequences of three identical inputs in a row, which could occur only once.

Given the trial-wise sequence of task inputs, we simulated beliefs of an agent that infers the probability of observing 1 according to a two–level HGF with a learning parameter $\omega_2 = -4$. Notably, this value is far from the Bayes-optimal value for the used sequence of inputs ($\omega_2 = -2.86$). We deliberately chose this sub-optimal learning parameter to demonstrate the model's ability to recover accurate estimates even when the underlying beliefs deviate significantly from the optimal HGF fit. This approach validates the model's robustness in capturing individual variability beyond what would be predicted by optimal Bayesian learning. The resulting simulated trial-wise belief trajectory was then used to generate behavioral outcomes according to different EAMs.

*Simulate behavioral outcomes*

For all models, we simulated four different task scenarios. In these scenarios, the intercepts of the EAMs parameters resulted in combinations of high (~90%) and low (~75%) accuracy, as well as fast (median ~ 350 ms) and slow (median ~ 700 ms) RTs. Furthermore, one parameter at a time was influenced by trial-wise beliefs to either a high (slopes of $\hat{\mu}$ equal to a maximum shift ±0.7 times the intercept of the associated parameter) or low extent (slopes of $\hat{\mu}$ equal to a maximum shift ±0.3 times the intercept of the associated parameter). Each simulation included 100 simulated participants.

Starting from the DDM, we conducted three sets of simulations in which either $w$, $a$, or $v$ were influenced by beliefs, respectively. The $w_0$ intercept was always .5, while the $a_a$ and $a_v$ intercepts for the four scenarios were, respectively: [$a_a = 1.2$; $a_v = 2$], [$a_a = 1.1$; $a_v = 0.97$], [$a_a = 2$; $a_v = 1.1$], [$a_a = 1.78$; $a_v = 0.62$]. For each set of simulations, the $b$ slope of $\hat{\mu}$ was adjusted to achieve a maximum shift in the intercept of the parameter influenced by trial-wise beliefs, either by ±0.3 or ±0.7 times, while it was set to 0 for the other two parameters. This resulted in eight different simulations for each set, for a total amount of 24 DDM simulations. In each simulation, trial-wise WFPT distributions for the lower and upper boundary were computed, following the model specifications described above. Next, trial-wise Ts and associated choices for the 100 simulated participants were randomly sampled from the combined lower and upper boundary WFPT distributions. Finally, a $Ter$ of 0.15 s was summed to T in order to obtain RT.

For the LNR, the $b_v$ and $b_i$ intercepts for the four scenarios were: [$a = -0.53$; $b_{val} = -0.47$], [$a = -0.75$; $b_{val} = -0.25$], [$a = 0.19$; $b_{val} = -0.49$], [$a = -0.10$; $b_{val} = -0.2$]. In combination with the two values of $b$, they resulted in eight different simulations. For all simulations, σ was set to 0.25. In each simulation, we computed trial-wise log-normal distributions for the two accumulators, following the model specifications described above. Then, for each simulated participant, a pair of RTs were sampled at each trial from the distributions of the two accumulators, and the lowest determined the RT and response choice for that trial. Since in our simulations the estimated $Ter$ parameter was highly correlated with all the LNR parameters (|ρ| > .9; see Supplementary Information 1 available at osf.io/pqc34), here we present simulations

in which *Ter* is assumed to be 0. Supplementary Information 1 presents analyses evaluating the inclusion of *Ter*.

For the RDM, we conducted two sets of simulations in which either *a* or *v* were influenced by beliefs, respectively. The $a_a$ and $a_v$ intercepts and the $b_{val}$ slope for the four scenarios were set to [$a_a$ = 2; $a_v$ =2.5; $b_{val}$ = *2.5*], [$a_a$ = 2; $a_v$ =3.55; $b_{val}$ = *1.45*], [$a_a$ = 3; $a_v$ =2; $b_{val}$ = *2*], [$a_a$ = 3; $a_v$ =2.84; $b_{val}$ = *1.16*]. These four scenarios, combined with the pair of values for the slopes $b_a$ and $b_v$ set as described for the DDM, generated 8 different simulations for each of the 2 sets, resulting in 16 simulations for each of the parameters.

**Model fitting**

The enhanced two-level HGF for binary inputs was first fitted using the trial-wise sequence of *u* as inputs only to obtain the Bayes optimal perceptual parameter $\omega_2$. Specifically, the model was configured as the default *tapas_ehgf_binary* except for the $\kappa$ parameter being fixed to 0 to unlink the third level from the rest of the model, and the prior variance of $\omega_3$ being set to 0 to avoid estimating it since it was excluded. The optimal $\omega_2$ value was -2.86 and was used as the prior mean for the analysis of simulated datasets.

Next, each simulated dataset was fitted using the *tapas_fitModel* procedure. Specifically, simulated RTs and choices were used as response inputs and *u* as task inputs. The perceptual model was configured as described above (i.e., $\kappa$ = 0; prior variance of $\omega_3$ = 0; prior mean of $\omega_2$ = -2.86).

The response model was one of the three EAMs according to the simulated dataset. Generally, parameter priors in the search space were specified to have a mean of 0 and a variance of 4 to have a large search space, with some exceptions.

In this context, each parameter's variance plays a crucial role in the optimization process, acting as an indicator of the spread or dispersion of possible parameter values. The parameter *Ter* was fixed to 0 for the LNR and RDM (but see Supplementary Information 1). Finally, on each dataset (excluding LNR), we carried out two analyses using full and reduced model configurations. Specifically, in the reduced model configuration, the slopes for $\hat{\mu}$ of the EAMs parameters not modulated by prior beliefs in the simulation were fixed to 0, and hence, excluded from the optimization process. In the full model configuration, we allowed the algorithm to optimize all slope parameters, including those related to parameters not modulated by the prior beliefs in the simulation.

**Results**

To evaluate the accuracy of the fitting procedure, we calculated the median recovered value and the 95% non-parametric confidence interval of each parameter for each parameter combination and model configuration. In the following sections, we will present the results for each EAM. Results will be summarized in tables for both model configurations. Moreover, histograms with overlaid Kernel density estimations will be provided in the main text for the first configuration and in Supplementary Information 2 (available at osf.io/pwh7e) for the second configuration. To facilitate the reading of the results, Table 1 summarizes the parameters estimated for both the perceptual and decision models.

**Table 1. PAM parameters.**

| Model | Parameter | Description |
|---|---|---|
| **Hierarchical Gaussian Filter** | $\omega_2 / \omega_3$ | Parameter influencing belief updating speed at the second and third HGF levels, respectively. |
| | $\kappa$ | Strength of the coupling between the second and the third HGF levels |
| **Drift Diffusion Model** | $b_w$ | Slope of the effect of prior beliefs on the starting point $w$. |
| | $a_a$ | Intercept of the boundary separation $a$. |
| | $b_a$ | Slope of the effect of precision of prior beliefs on the boundary separation $a$. |
| | $a_v$ | Intercept of the drift rate $v$. |
| | $b_v$ | Slope of the effect of prior beliefs on the drift rate $v$. |
| | $Ter$ | Non-decision time |
| **Lognormal Race Model** | $a$ | Intercept of lognormal mean. |
| | $b_{val}$ | Effect of the accumulator validity (i.e., the accumulator choice matches the stimulus) on the lognormal mean |
| | $b$ | Slope of the effect of prior beliefs on the lognormal mean. |
| | $\sigma^2$ | Lognormal variance. |
| **Racing Diffusion Model** | $a_a$ | Intercept of the decision threshold $a$. |
| | $b_a$ | Slope of the effect of prior beliefs on the decision threshold $a$. |
| | $a_v$ | Intercept of the drift rate $v$. |
| | $b_{val}$ | Effect of the accumulator validity (i.e., the accumulator choice matches the stimulus) on the drift rate $v$ |
| | $b_v$ | Slope of the effect of prior beliefs on the drift rate $v$. |

*Notes.* The table summarizes the parameters estimated in the different implementations of PAM along with their description.

*Results – DDM*

Results for the DDM are presented in Tables 2-4 and Figures 2-4. Specifically, Table 2 and Figure 2 present results of simulations in which $b_a$ and $b_v$ were simulated to be 0 (i.e., prior beliefs modulated only the starting point). Table 3 and Figure 3 present results of simulations in which $b_w$ and $b_v$ were simulated to be 0 (i.e., prior beliefs modulated only the boundary separation). Table 4 and Figure 4 present results of simulations in which $b_w$ and $b_a$ were simulated to be 0 (i.e., prior beliefs modulated only the drift rate).

Starting with the *a* parameter, median recovery values deviated by at most 1% from the simulated values. In most cases, differences between the reduced model and the full models were negligible, and in any case, they were always less than 1%. The 95% non-parametric confidence intervals always contained the true simulated values and were often larger in the full model compared to the reduced model.

Looking at the *v* parameter, median recovery values deviated by at most 2%. Differences between the reduced model and the full models were not observed. The 95% non-parametric confidence intervals always contained the true simulated values and differences between the two model configurations were negligible.

Concerning the slope parameters, their estimation was less accurate than that of the intercept parameters. Specifically, the median recovery values of $b_w$ deviated by at most 4% in the reduced model, while in the full model, they deviated by at most 10%. In the two scenarios in which $b_w$ was simulated to be 0, median recovery values in the full model deviated from 0 by at most ±0.02, and 0 was always included in the 95% non-parametric confidence intervals.

Similar results were observed for $b_a$. The median recovery values of $b_a$ deviated by at most 4% in both the reduced and the full models. In the two scenarios in which $b_a$ was simulated to be 0, median recovery values in the full model deviated from 0 by at most ±0.5, and 0 was always included in the 95% non-parametric confidence intervals.

The recovery of $b_v$ was less accurate and precise than that of $b_w$ and $b_a$. The median recovery values deviated by at most 13% in the reduced model, while in the full model, accuracy was higher, deviating up to 10%. In the two scenarios in which $b_v$ was simulated to be 0, median recovery values in the full model deviated from 0 by at most ±0.19, and 0 was always included in the 95% non-parametric confidence intervals.

The recovery of $Ter$ was very accurate, with median recovery values often equivalent to the simulated value or differing by at most 10 ms.

The recovery of $\omega_2$ was less accurate compared to the recovery of the parameters of the decision model. The median recovery values deviated by at most 28% in the reduced model and 14% in the full model. In all simulations, the median recovered value is recovered in the non-parametric confidence interval.

**Table 2. Parameter recovery for DDM simulated data with influence of prior beliefs on the starting point ($w$).**

| Simulated parameters | | | Estimated parameters | | | | | | | | | | |
|---|---|---|---|---|---|---|---|---|---|---|---|---|---|
| | | | Reduced model | | | | | Full model | | | | | |
| $a_a$ | $a_v$ | $b_w$ | $a_a$ | $a_v$ | $b_w$ | $T_{ER}$ | $\omega_2$ | $a_a$ | $a_v$ | $b_w$ | $b_a$ | $b_v$ | $T_{ER}$ | $\omega_2$ |
| 1.2 | 2 | 0.3 | 1.20 (0.06) | 2.04 (0.17) | 0.31 (0.1) | 0.15 (0.01) | -3.65 (1.32) | 1.20 (0.07) | 2.05 (0.17) | 0.33 (0.13) | 0.01 (0.28) | -0.18 (1.19) | 0.15 (0.01) | -3.70 (1.26) |

| 1.2 | 2 | 0.7 | 1.20 (0.05) | 2.00 (0.16) | 0.71 (0.08) | 0.15 (0.00) | -3.83 (0.63) | 1.20 (0.06) | 2.02 (0.16) | 0.70 (0.09) | -0.01 (0.26) | 0.19 (0.79) | 0.15 (0.00) | -3.81 (0.57) |
| --- | --- | --- | --- | --- | --- | --- | --- | --- | --- | --- | --- | --- | --- | --- |
| 1.1 | 0.97 | 0.3 | 1.10 (0.04) | 0.96 (0.10) | 0.31 (0.11) | 0.15 (0.00) | -3.75 (1.30) | 1.10 (0.06) | 0.96 (0.11) | 0.30 (0.11) | -0.01 (0.22) | 0.06 (0.78) | 0.15 (0.00) | -3.66 (1.32) |
| 1.1 | 0.97 | 0.7 | 1.10 (0.04) | 0.96 (0.12) | 0.69 (0.1) | 0.15 (0.00) | -3.82 (0.63) | 1.10 (0.05) | 0.95 (0.13) | 0.69 (0.12) | 0.03 (0.21) | 0.11 (0.75) | 0.15 (0.00) | -3.83 (0.62) |
| 2 | 1.1 | 0.3 | 1.98 (0.08) | 1.13 (0.10) | 0.31 (0.08) | 0.15 (0.02) | -3.76 (1.11) | 1.98 (0.13) | 1.13 (0.1) | 0.32 (0.11) | 0.00 (0.37) | 0.07 (0.39) | 0.15 (0.02) | -3.85 (1.2) |
| 2 | 1.1 | 0.7 | 1.97 (0.08) | 1.12 (0.09) | 0.68 (0.1) | 0.15 (0.01) | -3.88 (0.54) | 1.98 (0.09) | 1.13 (0.09) | 0.69 (0.13) | 0.01 (0.47) | 0.08 (0.55) | 0.15 (0.01) | -3.89 (0.54) |
| 1.78 | 0.62 | 0.3 | 1.75 (0.06) | 0.62 (0.07) | 0.31 (0.1) | 0.15 (0.01) | -3.56 (1.16) | 1.76 (0.09) | 0.62 (0.08) | 0.30 (0.12) | -0.01 (0.33) | -0.04 (0.51) | 0.15 (0.01) | -3.66 (1.47) |
| 1.78 | 0.62 | 0.7 | 1.76 (0.06) | 0.64 (0.08) | 0.71 (0.09) | 0.15 (0.01) | -3.91 (0.52) | 1.75 (0.06) | 0.64 (0.08) | 0.70 (0.12) | 0.01 (0.29) | 0.14 (0.46) | 0.15 (0.01) | -3.93 (0.53) |

Notes: In this simulation parameters $b_a$ and $b_v$ were simulated to be 0, $T_{ER}$ to be 0.15 s. and $\omega_2$ to be -4. Values indicate median recovery parameters and the associated interquartile range.

**Table 3. Parameter recovery for DDM simulated data with influence of prior beliefs on the boundary separation (*a*).**

| Simulated parameters | | | Estimated parameters | | | | | | | | | | | |
| --- | --- | --- | --- | --- | --- | --- | --- | --- | --- | --- | --- | --- | --- | --- |
| | | | Reduced model | | | | | Full model | | | | | | |
| $a_a$ | $a_v$ | $b_a$ | $a_a$ | $a_v$ | $b_a$ | $T_{ER}$ | $\omega_2$ | $a_a$ | $a_v$ | $b_w$ | $b_a$ | $b_v$ | $T_{ER}$ | $\omega_2$ |
| 1.2 | 2 | -0.72 | 1.20 (0.05) | 2.01 (0.16) | -0.74 (0.21) | 0.15 (0.00) | -3.78 (0.81) | 1.20 (0.05) | 2.01 (0.16) | -0.01 (0.10) | -0.72 (0.20) | 0.11 (0.86) | 0.15 (0.00) | -3.80 (0.79) |
| 1.2 | 2 | -1.68 | 1.20 (0.05) | 2.01 (0.2) | -1.65 (0.15) | 0.15 (0.00) | -3.91 (0.36) | 1.20 (0.05) | 2.03 (0.23) | 0.00 (0.11) | -1.66 (0.14) | -0.06 (1.22) | 0.15 (0.00) | -3.90 (0.35) |
| 1.1 | 0.97 | -0.66 | 1.10 (0.05) | 0.95 (0.16) | -0.67 (0.17) | 0.15 (0.00) | -3.80 (0.85) | 1.10 (0.05) | 0.97 (0.16) | 0.00 (0.11) | -0.68 (0.18) | 0.00 (0.86) | 0.15 (0.00) | -3.77 (0.93) |
| 1.1 | 0.97 | -1.54 | 1.09 (0.04) | 0.97 (0.18) | -1.51 (0.12) | 0.15 (0.00) | -3.89 (0.33) | 1.09 (0.04) | 0.97 (0.16) | 0.01 (0.09) | -1.51 (0.12) | -0.03 (1.20) | 0.15 (0.00) | -3.89 (0.33) |
| 2 | 1.1 | -1.2 | 1.98 (0.08) | 1.12 (0.11) | -1.17 (0.30) | 0.15 (0.01) | -3.77 (0.89) | 1.98 (0.08) | 1.12 (0.11) | -0.02 (0.13) | -1.16 (0.32) | -0.14 (0.65) | 0.16 (0.01) | -3.74 (0.95) |
| 2 | 1.1 | -2.8 | 1.95 (0.07) | 1.10 (0.09) | -2.69 (0.21) | 0.15 (0.00) | -3.87 (0.46) | 1.95 (0.07) | 1.11 (0.08) | 0.00 (0.09) | -2.69 (0.21) | -0.05 (0.66) | 0.15 (0.00) | -3.85 (0.45) |
| 1.78 | 0.62 | -1.06 | 1.76 (0.08) | 0.64 (0.09) | -1.05 (0.30) | 0.15 (0.01) | -3.8 (0.83) | 1.76 (0.07) | 0.64 (0.08) | 0.00 (0.13) | -1.04 (0.30) | 0.04 (0.60) | 0.15 (0.01) | -3.75 (0.75) |
| 1.78 | 0.62 | -2.49 | 1.74 (0.07) | 0.62 (0.10) | -2.39 (0.20) | 0.15 (0.00) | -3.86 (0.30) | 1.74 (0.06) | 0.62 (0.10) | 0.00 (0.12) | -2.39 (0.20) | 0.11 (0.87) | 0.15 (0.00) | -3.88 (0.30) |

Notes: In this simulation parameters $b_w$ and $b_v$ were simulated to be 0, $T_{ER}$ to be 0.15 s, and $\omega_2$ to be -4. Values indicate median recovery parameters and the associated interquartile range.

**Table 4. Parameter recovery for DDM simulated data with influence of prior beliefs on the drift rate ($v$).**

| Simulated parameters | | | Estimated parameters | | | | | | | | | | |
|---|---|---|---|---|---|---|---|---|---|---|---|---|---|
| | | | Reduced model | | | | | Full model | | | | | |
| $a_a$ | $a_v$ | $b_v$ | $a_a$ | $a_v$ | $b_v$ | $T_{ER}$ | $\omega_2$ | $a_a$ | $a_v$ | $b_w$ | $b_a$ | $b_v$ | $T_{ER}$ | $\omega_2$ |
| 1.20 | 2.00 | 1.20 | 1.21 (0.06) | 2.01 (0.11) | 1.02 (0.78) | 0.15 (0.01) | -3.14 (1.17) | 1.21 (0.09) | 2.01 (0.12) | 0.02 (0.12) | -0.04 (0.32) | 1.01 (0.88) | 0.15 (0.01) | -3.45 (1.64) |
| 1.20 | 2.00 | 2.80 | 1.20 (0.05) | 2.00 (0.12) | 2.6 (0.73) | 0.15 (0.01) | -3.66 (0.77) | 1.20 (0.05) | 2.00 (0.12) | 0.03 (0.14) | 0.01 (0.32) | 2.49 (1.01) | 0.15 (0.01) | -3.67 (0.85) |
| 1.10 | 0.97 | 0.58 | 1.10 (0.04) | 0.99 (0.11) | 0.59 (0.74) | 0.15 (0.00) | -2.89 (0.81) | 1.10 (0.06) | 0.99 (0.11) | 0.00 (0.12) | -0.03 (0.30) | 0.55 (0.88) | 0.15 (0.00) | -3.15 (1.79) |
| 1.10 | 0.97 | 1.36 | 1.10 (0.04) | 0.96 (0.15) | 1.27 (0.71) | 0.15 (0.01) | -3.31 (1.42) | 1.01 (0.05) | 0.96 (0.16) | 0.03 (0.12) | 0.04 (0.27) | 1.14 (0.93) | 0.15 (0.01) | -3.26 (1.79) |
| 2.00 | 1.10 | 0.66 | 1.97 (0.09) | 1.11 (0.09) | 0.68 (0.43) | 0.15 (0.02) | -3.27 (1.15) | 1.98 (0.10) | 1.12 (0.09) | 0.00 (0.13) | -0.08 (0.50) | 0.66 (0.55) | 0.16 (0.02) | -3.31 (1.40) |
| 2.00 | 1.10 | 1.54 | 1.98 (0.07) | 1.09 (0.10) | 1.46 (0.56) | 0.16 (0.01) | -3.76 (1.04) | 1.98 (0.13) | 1.09 (0.10) | 0.03 (0.17) | -0.02 (0.60) | 1.41 (0.58) | 0.16 (0.02) | -3.80 (1.30) |
| 1.78 | 0.62 | 0.37 | 1.75 (0.07) | 0.63 (0.09) | 0.33 (0.52) | 0.15 (0.02) | -3.05 (0.92) | 1.76 (0.08) | 0.63 (0.09) | -0.01 (0.15) | -0.09 (0.45) | 0.31 (0.68) | 0.16 (0.01) | -3.50 (1.55) |
| 1.78 | 0.62 | 0.87 | 1.76 (0.06) | 0.64 (0.10) | 0.75 (0.63) | 0.15 (0.01) | -3.36 (1.06) | 1.75 (0.09) | 0.65 (0.10) | 0.02 (0.14) | -0.04 (0.51) | 0.69 (0.85) | 0.16 (0.01) | -3.51 (1.87) |

Notes: In this simulation parameters $b_w$ and $b_a$ were simulated to be 0, $T_{ER}$ to be .15 s, and $\omega_2$ to be -4. Values indicate median recovery parameters and the associated interquartile range.

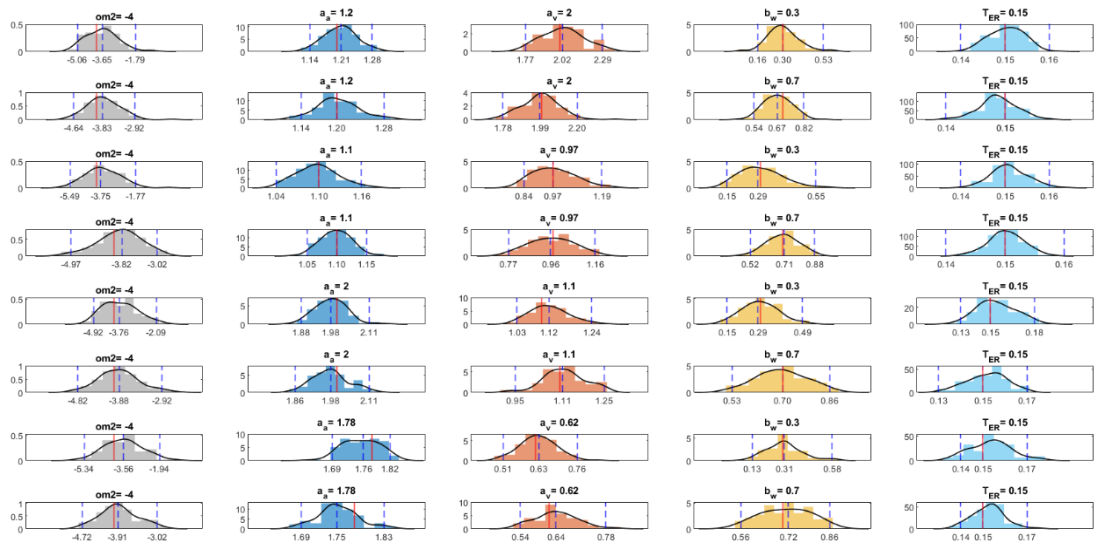

**Figure 2. Parameter recovery of the DDM reduced model for simulations in which the starting point *w* was modulated by prior beliefs.** The figure shows histograms with Kernel Density Estimation for the four parameters of the reduced DDM recovered across different simulated conditions. Each column represents one of the four parameters, with the true simulated parameter reported as the title at the top of each column, while each row corresponds to a unique combination of simulated parameters. The red vertical line in each plot indicates the true simulated parameter value, while the blue dotted lines represent the 2.5th, 50th (median), and 97.5th percentiles of the estimated parameters

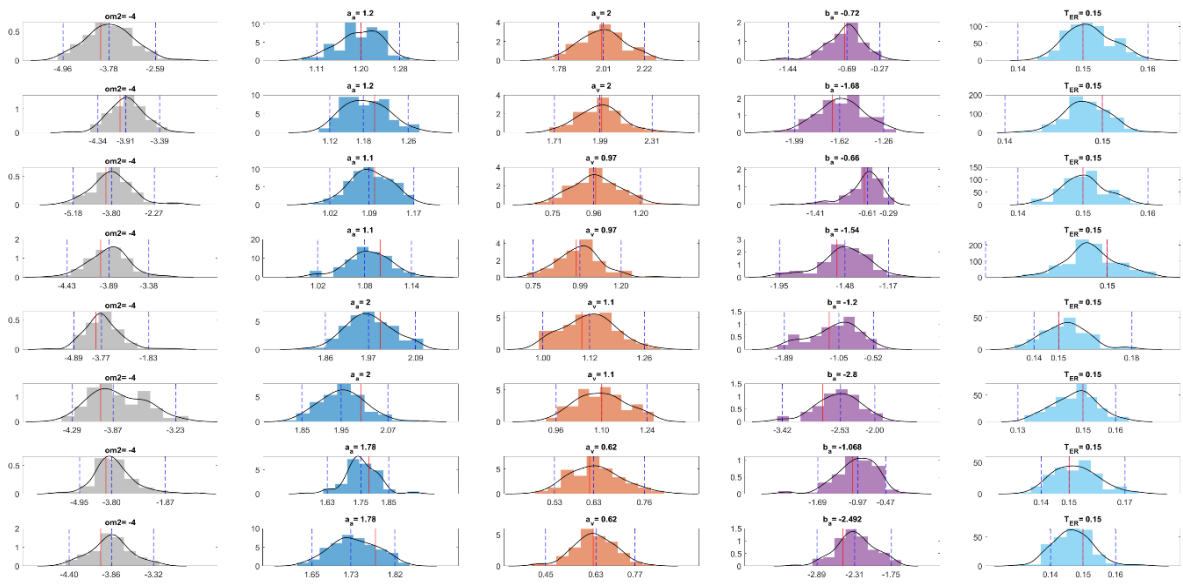

**Figure 3. Parameter recovery of the DDM reduced model for simulations in which the boundary separation *a* was modulated by prior beliefs.** For conventions see Figure 2.

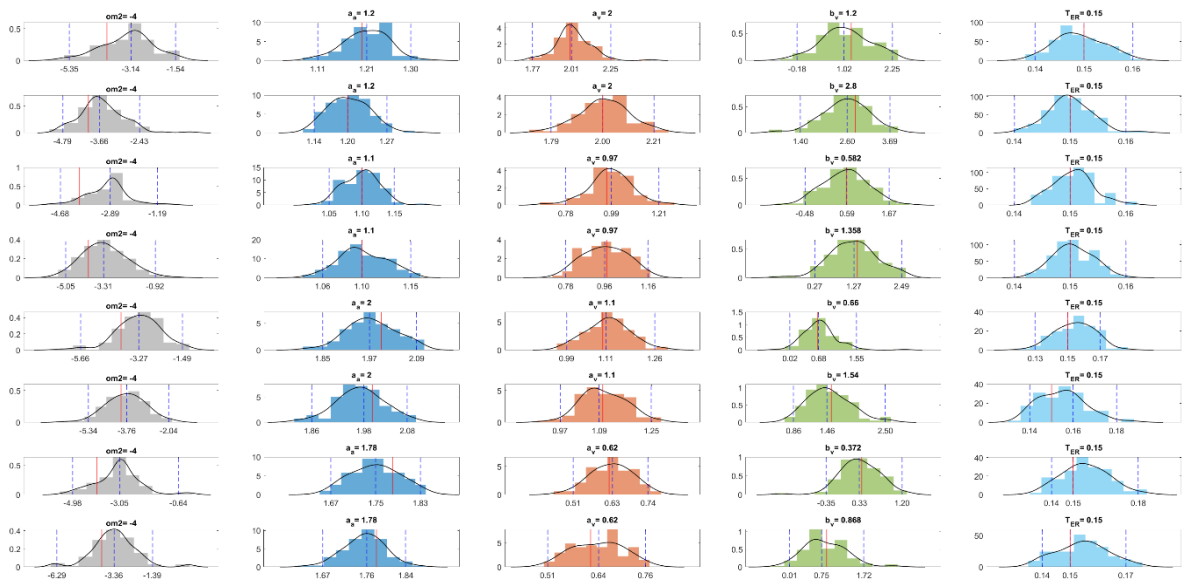

**Figure 4. Parameter recovery of the DDM reduced model for simulations in which the drift rate *v* was modulated by prior beliefs.** For conventions see Figure 2.

*Results – LNR*

Results for the LNR are presented in Table 5 and Figure 5. The median recovery values for the *a* and $b_{val}$ intercepts were almost equivalent to simulated values. The 95% non-parametric confidence intervals for both *a* and $b_{val}$ always contained the true simulated value. The recovery of the slope parameter *b* was less accurate (deviating by at most 15%). The median recovery values σ were equivalent to the simulated value. The recovery of $\omega_2$ was generally accurate with a maximum deviation from the true parameter of 10%.

**Table 5. Parameter recovery for LNR simulated data.**

| Simulated Parameters | | | | Estimated Parameters | | | | |
|---|---|---|---|---|---|---|---|---|
| a | $b_{val}$ | b | σ | a | $b_{val}$ | b | σ | ω2 |
| -0.53 | -0.47 | -1.04 | 0.25 | -0.53 (0.03) | -0.47 (0.03) | -0.63 (0.12) | 0.25 (0.01) | -3.92 (0.64) |
| -0.53 | -0.47 | -0.19 | 0.25 | 0.28 (-0.53) | 0.04 (-0.47) | 0.04 (-1.41) | 0.25 (0.01) | -3.96 (0.28) |
| -0.75 | -0.25 | -1.26 | 0.25 | 0.50 (-0.75) | 0.03 (-0.25) | 0.04 (-0.61) | 0.25 (0.01) | -3.95 (0.5) |
| -0.75 | -0.25 | -0.41 | 0.25 | 0.28 (-0.75) | 0.03 (-0.25) | 0.04 (-1.42) | 0.25 (0.01) | -3.98 (0.28) |
| 0.19 | -0.49 | -0.32 | 0.25 | 0.67 (0.19) | 0.04 (-0.49) | 0.05 (-0.61) | 0.25 (0.01) | -3.97 (0.67) |
| 0.19 | -0.49 | 0.53 | 0.25 | 0.22 (0.19) | 0.04 (-0.50) | 0.04 (-1.43) | 0.25 (0.01) | -3.93 (0.22) |
| -0.1 | -0.2 | -0.61 | 0.25 | 0.59 (-0.10) | 0.02 (-0.20) | 0.03 (-0.61) | 0.25 (0.01) | -3.93 (0.59) |
| -0.1 | -0.2 | 0.24 | 0.25 | 0.28 (-0.10) | 0.03 (-0.20) | 0.04 (-1.45) | 0.25 (0.01) | -4.01 (0.28) |

Notes: In this simulation the parameter $T_{ER}$ was simulated to be 0 and $\omega_2$ to be -4. Values indicate median recovery parameters and the associated interquartile range.

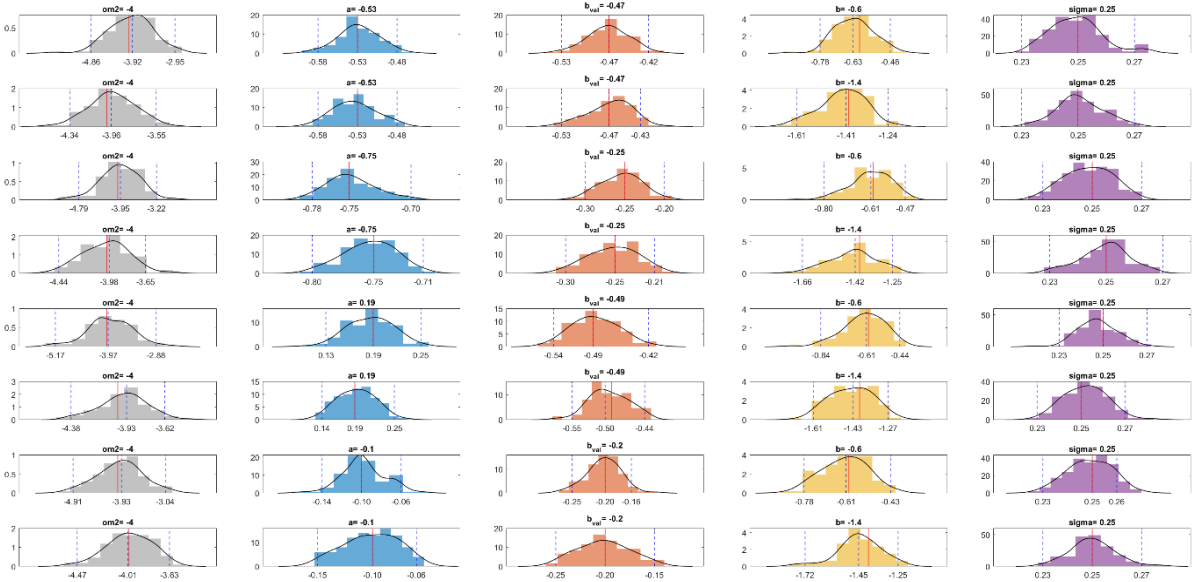

**Figure 5. Parameter recovery of the LNR model.** For conventions see Figure 2.

*Results – RDM*

Results for the RDM are presented in Tables 6-7 and Figures 6-7. Similar to the LNR, also recovery of RDM parameters was highly accurate and precise. Specifically, Table 6 and Figure 6 present results of simulations in which $b_v$ was simulated to be 0 (i.e., prior beliefs modulated only the decision boundary), while Table 7 and Figure 7 present results of simulations in which $b_a$ was simulated to be 0 (i.e., prior beliefs modulated only the only the drift rate).

The median recovery values for the $a_a$, $a_v$, and $b_{val}$ intercepts deviated by at most 2%. In most cases, differences between the reduced models and the full models were negligible, and in all cases, they were always less than 2%. The 95% non-parametric confidence intervals always contained the true simulated values and were similar between the two model configurations.

As observed for the DDM and LNR, the recovery of the slope parameters $b_a$ and $b_v$ was less accurate (deviating by at most 13% in the reduced model). Generally, the performance of the full models was worse in terms of accuracy. Finally, in the scenarios in which $b_a$ or $b_v$ were simulated to be 0, median recovery values for $b_a$ in the full model deviated from 0 by at most ±0.6, while median recovery values for $b_v$ deviated from 0 by at most ±0.7. In all cases 0 was always included in the 95% non-parametric confidence intervals. The recovery of $\omega_2$ was generally accurate. The median recovery values deviated by at most 14% from the true parameter for the reduced model and by 4% for the full model.

**Table 6. Parameter recovery for RDM simulated data with influence of prior beliefs on the boundary (*a*).**

| Simulated Parameters | | | | Estimated Parameters | | | | | | | | | | |
|---|---|---|---|---|---|---|---|---|---|---|---|---|---|---|
| | | | | Reduced Model | | | | | Full Model | | | | | |
| $a_a$ | $a_v$ | $b_{val}$ | $b_a$ | $a_a$ | $a_v$ | $b_{val}$ | $b_a$ | $\omega_2$ | $a_a$ | $a_v$ | $b_{val}$ | $b_a$ | $b_v$ | $\omega_2$ |

| | | | | | | | | | | | | | |
|---|---|---|---|---|---|---|---|---|---|---|---|---|---|
| 2 | 2.5 | 2.5 | -1.2 | 2.00 (0.10) | 2.50 (0.29) | 2.51 (0.27) | -1.22 (0.27) | -3.85 (0.74) | 1.99 (0.10) | 2.50 (0.32) | 2.52 (0.27) | -1.15 (0.53) | 0.16 (1.27) | -3.83 (0.78) |
| 2 | 2.5 | 2.5 | -2.8 | 1.98 (0.08) | 2.47 (0.28) | 2.51 (0.24) | -2.80 (0.28) | -3.95 (0.34) | 1.97 (0.09) | 2.42 (0.31) | 2.53 (0.25) | -2.74 (0.4) | 0.29 (0.88) | -3.95 (0.34) |
| 2 | 3.55 | 1.45 | -1.2 | 2.00 (0.09) | 3.54 (0.32) | 1.45 (0.2) | -1.21 (0.25) | -3.89 (0.62) | 2.00 (0.10) | 3.53 (0.32) | 1.46 (0.21) | -1.18 (0.5) | -0.02 (1.12) | -3.91 (0.57) |
| 2 | 3.55 | 1.45 | -2.8 | 1.99 (0.10) | 3.48 (0.30) | 1.48 (0.25) | -2.81 (0.27) | -3.97 (0.25) | 1.98 (0.12) | 3.48 (0.34) | 1.49 (0.25) | -2.72 (0.5) | 0.28 (1.29) | -3.96 (0.26) |
| 3 | 2 | 2 | -1.8 | 3.02 (0.13) | 2.02 (0.21) | 1.99 (0.17) | -1.82 (0.39) | -3.91 (0.64) | 3.00 (0.15) | 2.01 (0.25) | 2.00 (0.19) | -1.66 (0.62) | 0.28 (0.85) | -3.93 (0.65) |
| 3 | 2 | 2 | -4.2 | 2.97 (0.13) | 2.01 (0.24) | 1.97 (0.20) | -4.21 (0.47) | -3.93 (0.28) | 2.94 (0.13) | 1.97 (0.23) | 1.98 (0.19) | -3.96 (0.66) | 0.44 (1.01) | -3.92 (0.29) |
| 3 | 2.84 | 1.16 | -1.8 | 3.00 (0.16) | 2.82 (0.23) | 1.16 (0.16) | -1.83 (0.40) | -3.88 (0.57) | 2.98 (0.16) | 2.80 (0.26) | 1.16 (0.16) | -1.71 (0.77) | 0.17 (1.07) | -3.87 (0.56) |
| 3 | 2.84 | 1.16 | -4.2 | 2.97 (0.16) | 2.80 (0.22) | 1.17 (0.15) | -4.15 (0.42) | -3.93 (0.27) | 2.91 (0.15) | 2.72 (0.23) | 1.18 (0.16) | -3.74 (0.61) | 0.75 (0.96) | -3.94 (0.26) |

Notes: In this simulation parameters $b_v$ and $T_{ER}$ were simulated to be 0 and $\omega_2$ was simulated to be -4. Values indicate median recovery parameters and the associated interquartile range.

**Table 7. Parameter recovery for RDM simulated data with influence of prior beliefs on the boundary (v).**

| Simulated Parameters | | | | Estimated Parameters | | | | | | | | | |
|---|---|---|---|---|---|---|---|---|---|---|---|---|---|
| | | | | Reduced Model | | | | | Full Model | | | | |
| $a_v$ | $b_{val}$ | $b_v$ | | $a_a$ | $a_v$ | $b_{val}$ | bv | $\omega_2$ | $a_a$ | $a_v$ | $b_{val}$ | $b_a$ | $b_v$ | $\omega_2$ |
| -4 | 2 | 2.5 | 2.5 | 2.00 (0.09) | 2.50 (0.28) | 2.52 (0.20) | -1.43 (0.68) | -3.61 (1.31) | 2.01 (0.08) | 2.51 (0.28) | 2.51 (0.20) | 0.19 (0.49) | -1.22 (1.00) | -3.57 (1.41) |
| -4 | 2 | 2.5 | 2.5 | 2.01 (0.08) | 2.52 (0.26) | 2.50 (0.29) | -3.32 (0.66) | -3.74 (0.63) | 2.02 (0.08) | 2.53 (0.25) | 2.49 (0.29) | 0.25 (0.40) | -2.86 (0.94) | -3.74 (0.66) |
| -4 | 2 | 3.55 | 1.45 | 2.00 (0.09) | 3.57 (0.27) | 1.45 (0.16) | -2.02 (0.57) | -3.69 (0.91) | 2.01 (0.09) | 3.58 (0.26) | 1.44 (0.17) | 0.17 (0.43) | -1.83 (1.21) | -3.72 (1.01) |
| -4 | 2 | 3.55 | 1.45 | 2.00 (0.08) | 3.63 (0.25) | 1.39 (0.17) | -4.75 (0.58) | -3.87 (0.49) | 2.02 (0.09) | 3.68 (0.27) | 1.38 (0.17) | 0.39 (0.39) | -3.81 (0.98) | -3.92 (0.50) |
| -4 | 3 | 2 | 2 | 3.02 (0.16) | 2.05 (0.22) | 1.97 (0.18) | -1.15 (0.41) | -3.7 (0.92) | 3.02 (0.16) | 2.05 (0.22) | 1.97 (0.18) | 0.13 (0.69) | -0.98 (0.98) | -3.59 (0.93) |
| -4 | 3 | 2 | 2 | 3.00 (0.14) | 2.00 (0.20) | 1.98 (0.20) | -2.79 (0.54) | -3.86 (0.59) | 3.01 (0.14) | 2.01 (0.20) | 1.98 (0.20) | 0.24 (0.77) | -2.54 (0.91) | -3.87 (0.60) |
| -4 | 3 | 2.84 | 1.16 | 2.99 (0.10) | 2.81 (0.20) | 1.19 (0.15) | -1.69 (0.52) | -3.78 (0.79) | 3.00 (0.11) | 2.82 (0.19) | 1.18 (0.15) | 0.15 (0.63) | -1.56 (0.90) | -3.79 (0.78) |
| -4 | 3 | 2.84 | 1.16 | 2.96 (0.13) | 2.83 (0.20) | 1.11 (0.15) | -3.9 (0.45) | -3.92 (0.40) | 3.02 (0.15) | 2.89 (0.21) | 1.11 (0.16) | 0.59 (0.82) | -3.15 (1.16) | -3.96 (0.40) |

Notes: In this simulation parameters $b_a$ and $T_{ER}$ were simulated to be 0 and $\omega_2$ was simulated to be -4. Values indicate median recovery parameters and the associated interquartile range.

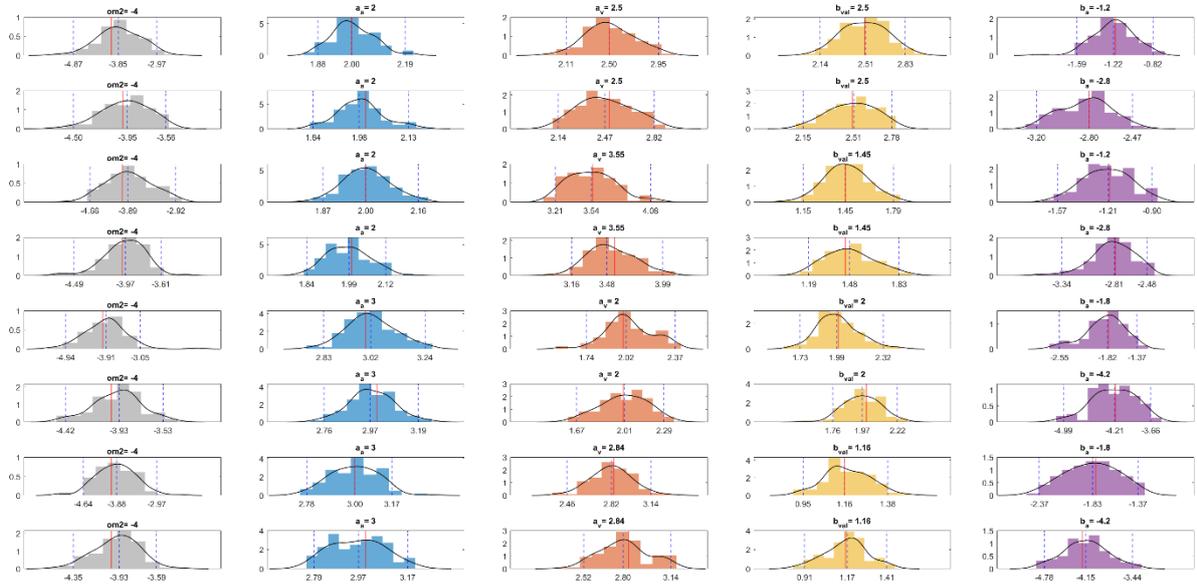

**Figure 6. Parameter recovery of the RDM reduced model for simulations in which the decision boundary *a* was modulated by prior beliefs.** For conventions see Figure 2.

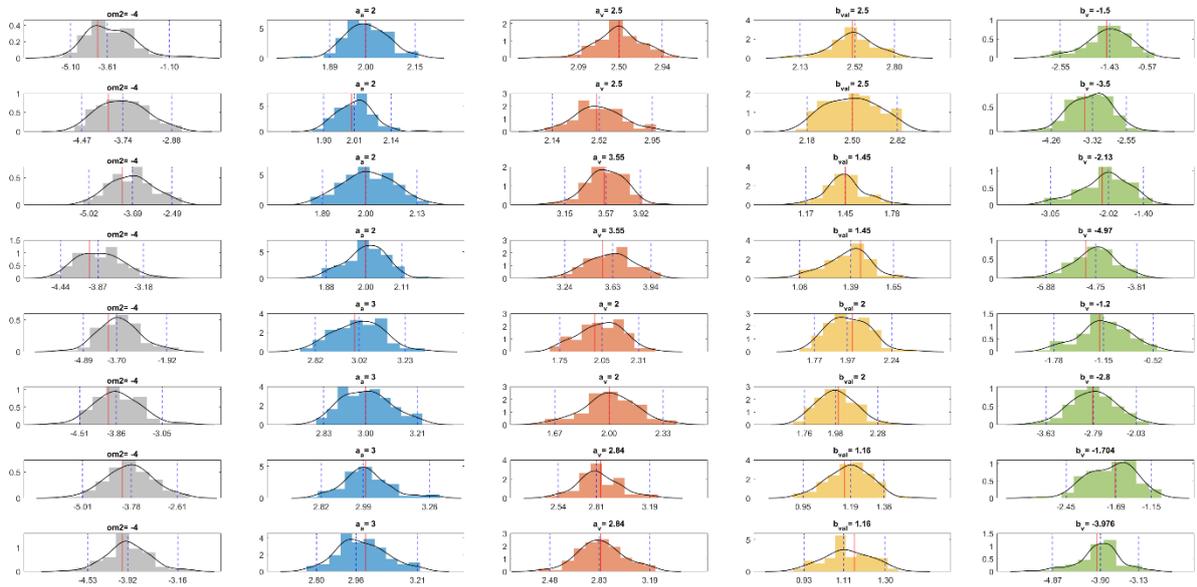

**Figure 7. Parameter recovery of the RDM reduced model for simulations in which the drift rate *v* was modulated by prior beliefs.** For conventions see Figure 2.

**STEP-BY-STEP TUTORIAL**

In this section, we provide a MATLAB tutorial on how to use PAM with a real dataset. In the first part, we outline the perceptual decision-making study that led to the dataset. Next, we describe all the PAM steps, from data preprocessing to statistical inference. A MATLAB live-script of the tutorial is available at [osf.io/3jve9/](osf.io/3jve9/). Furthermore, several PAM examples with different perceptual and decision models, as well as using different programming languages are available at [github.com/antovis86/PAM-PredictiveAccumulationModels](github.com/antovis86/PAM-PredictiveAccumulationModels).

**Experiment description.**

*Participants*

Forty participants completed the experimental task (22 females, 18 males; mean age = 21.8 years, SD = 1.5). All participants reported having normal or corrected-to-normal visual acuity, no current or past neurological or psychiatric disorders, and not being under the influence of alcohol or other drugs that might affect cognitive functioning. Participants gave their informed consent to participate in the study, which was conducted in accordance with the ethics standards of the 2013 Declaration of Helsinki for human studies of the World Medical Association. The study was previously approved by the Ethical Committee for the Psychological Research of the University of Padova (approved protocol reference number: 4439).

*Task and procedure*

The visual stimuli were random dot kinematograms (RDKs). Each RDK comprised 600 dots uniformly pseudo-randomly distributed within a circular area of 4° (density: 10 dots/deg²), centered on a 23.6-inch screen (1920 × 1080 pixels; refresh rate: 60 Hz). Dots were visible only within an aperture of 2.5° radius, centered on the screen. When visible, the dots were black against a pale cornflower blue (RGB: 189, 215, 238) background. Each dot, with a diameter of 0.05°, moved at a constant speed of 18°/s. Dots were categorized as either signal or noise dots. Signal dots consistently moved in the global motion direction of the RDK (left or right), while noise dots moved in all other directions. When a dot exited the 4° circular area, it reentered from the opposite side.

Each trial (Figure 8A) started with the presentation of a red fixation dot (diameter: 0.15°) displayed at the center of the screen until the end of the trial. After 500 ms, an RDK was displayed for 6 frames. Participants were asked to judge the leftward or rightward direction of coherent dots by pressing the "F" and "J" keys with their left and right index fingers, respectively (the maximum allowed time to respond was 1500 ms). The trial ended 1500 ms after the RDK offset.

The task was divided into an adaptive phase and a subsequent testing phase (Figure 8B). The adaptive phase consisted of 200 trials of a psychophysical Bayesian adaptive procedure to estimate the individual level of RDK coherence (i.e., the proportion of signal dots) required to achieve 90% accuracy. Specifically, we used Luigi Acerbi's MATLAB implementation of the PSI method (https://github.com/lacerbi/psybayes) by Kontsevich and Tyler (1999), extended to include the lapse rate (Prins, 2012). In the adaptive phase, the probability of observing a rightward movement was 0.5. The testing phase consisted of 400 trials in which RDK coherence corresponded to the level estimated in the adaptive phase, while the direction was as described in the Data Simulation section. Thirty-second breaks were provided between the adaptive and testing phases, as well as every three blocks of the testing phase.

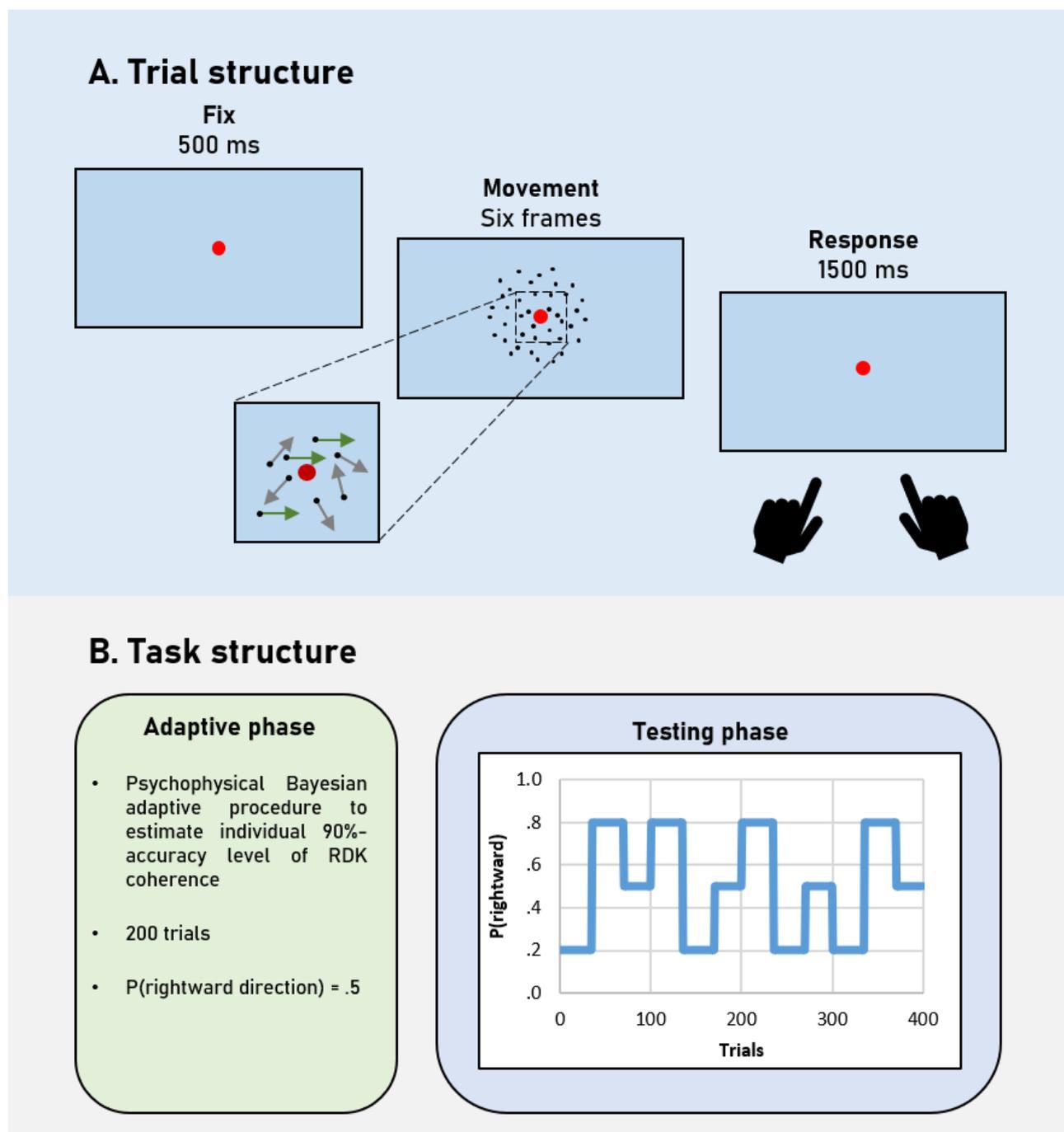

**Figure 8. Random Dot Kinematograms (RDK) task.** (A) Example of a trial: A central red fixation dot appeared and remained onscreen throughout the trial. After 500 ms, a RDK was presented for 6 frames (~100 ms). Participants indicated perceived coherent dot motion direction (left or right) using "F" and "J" keys, respectively, with a 1500 ms response window. (B) The experiment consisted of two phases. The adaptive phase (200 trials) used a Bayesian adaptive procedure to estimate individual RDK coherence thresholds for 90% accuracy, with rightward motion probability fixed at 0.5. The subsequent testing phase (400 trials) employed the RDK coherence level determined in the adaptive phase, while the probability of rightward motion varied as indicated by the solid blue line.

## MATLAB tutorial

In this section we provide a narrative description of steps of the PAM procedure implemented in the tutorial live-script (osf.io/t9rqn).

*Data preprocessing*

In this step, participants' data from the RDK testing phase are preprocessed into the correct format for the model fitting step. Specifically, the trial-wise sequence of task inputs is converted into a vector *u* of zeros and ones, corresponding respectively to the leftward or rightward trial-wise RDK direction. For each participant, a two-column matrix *y* (with the number of rows equal to the length of *u*) is built by concatenating trial-wise RTs (first column) and response choices (second column). Concerning the RT column, rows including missing or anticipated (RT < 0.15s) responses are marked as *NaN*. Moreover, RTs are expressed in seconds for all three models. Response choices are coded as *u*, using zeros and ones for leftward or rightward responses, respectively.

*Model configuration*

This step preliminarily requires configuring perceptual and decision models. Regarding the perceptual model, the *tapas_ehgf_binary* model was configured to have two levels, and the prior mean for $\omega_2$ was set to the Bayes-optimal value for our sequence of task events (see the Parameter Recovery section above). The decision to use a two-level HGF (instead of the default three-level HGF) was based on the task structure. Specifically, due to the regularity of probability changes during the task, we assumed that estimations of environmental volatility (third HGF level) would have negligible influence on the lower levels, making the two-level HGF an appropriate and more parsimonious model.

Regarding decision models, to provide a more comprehensive tutorial, we decided to employ eight different PAM models. Specifically, we used three different DDM configurations that separately assess the influence of predictions on the *w*, *a*, and *v* parameters (the models are labeled: DDM_w, DDM_a, and DDM_v), as well as the default configuration assessing all prediction effects on all parameters (DDM_full). Similarly, we used two RDM configurations, one for prediction effects on *a* and one for *v* (RDM_a and RDM_v) and the default configuration assessing prediction effects on both parameters (RDM_full). The configurations of DDM_w, DDM_a, and DDM_v, RDM_a and RDM_v require setting the prior mean and variance of the *b* slopes associated with the other parameter(s) to 0. Concerning the LNR, we used the default configuration. Of note, based on the simulation results, the *Ter* parameters for the RDM and LNR were fixed to 0, which is the default configuration. However, *Ter* can be configured to be estimated by setting its prior variance to a positive value.

*Model fitting*

Once the *u* and *y* inputs are prepared and the model configurations specified, the *tapas_fitModel* function can be used to determine the maximum a posteriori estimates of individual parameters for both the perceptual and decision models through the BFGS quasi-Newton optimization algorithm. This procedure is performed for every combination of selected perceptual and decision models. Specifically, here for each participant, the estimation of model parameters was conducted eight times, once for each decision-model configuration.

*Statistical inference on estimated parameters.*

Perceptual and decision parameters estimated at the individual level can be used for statistical inference at the group level. Specifically, in the present experiment, we assessed whether prior beliefs about motion direction influenced EAM parameters by performing two-tailed one-sample t-tests, separately for *b* slopes of each decision model configuration. As mentioned above, this can be done since *b* slopes in PAM models are allowed to have bidirectional effects on the EAM parameters. Notably, other statistical tests can be conducted on the basis of the specific experiment on all perceptual and decision parameters. For example, paired-sample t-tests or repeated measures ANOVAs can assess differences in two or more experimental situations, while two-sample t-tests or ANOVAs can evaluate between-group differences. Furthermore, parameters can be correlated with other individual measures.

*Bayesian Model selection*

Bayesian Model Selection (BMS) is a valid method used to compare the relative log-evidence of different models (LME). LME is a measure of model goodness calculated as the negative surprise about the data given the specified model, which considers both model accuracy and complexity (Rigoux et al., 2014; Stephan et al., 2009). Since the *tapas_fitModel* function returns LME (along with two other popular measures of model goodness, namely AIC and BIC), BMS can be used for model comparison. Specifically here, random effects BMS, as implemented in the *spm_BMS_gibbs* function from the SPM software package (http://www.fil.ion.ucl.ac.uk/spm/) of the SPM toolbox, was used to evaluate the relative plausibility of the different decision model configurations. For the DDM and RDM, BMS was first performed among their variants. Next, we compared the best DDM, the best RDM and the LNR. BMS results are reported in terms of exceedance probabilities (Stephan et al., 2009). Exceedance probability is a Bayesian measure that helps determine which model best explains a group's data by estimating the likelihood that one model outperforms others across participants.

**Experiment results**

The BMS among the three DDM configurations revealed that the model in which beliefs modulated *v* (DDM_v) had clearly higher model evidence (excedance probability, $xp = .9991$) than the DDM_w *($xp = .0009$)*, DDM_a *($xp < .0001$)*, or the DDM_full *($xp < .0001$)*. The statistical assessment of the impact of prior beliefs on *v* in the DDM_v model showed a significant positive effect ($t(39) = 9.62$, $p < .001$), indicating a higher drift rate with higher prior probability of the observed motion direction.

Concerning the LNR, the effect of prior beliefs was significant ($t(39) = -10.14$, $p < .001$).

Looking at the RDM, the BMS showed that the model assuming a belief modulation of $v$ (RDM_v) had higher evidence ($xp = .9988$) than RDM_a ($xp = .0012$) and RDM_full ($xp < .0001$). The statistical assessment of the impact of prior beliefs on $v$ in the RDM_v model showed a significant positive effect ($t(39) = 10.24$, $p < .001$), indicating a higher drift rate with higher prior probability of the observed motion direction. Finally, BMS among DDM_v, LNR, and RDM_v, showed that the winning model was the RDM_v (exceedance probabilities: DDM_v $= .0221$, LNR $< .0001$, RDM_v $= .9779$).

Concerning computational time, we report the median fitting time for the three full EAMs as from analyses performed on a Dell G15 5530, an x64-based PC with an Intel64 Family 6 Model 183 processor at 2600 MHz and 32453 MB of RAM. The median fitting time for the full DDM was 19 seconds (IQR $= 10$ seconds), and the total time for fitting the 40 datasets without parallelization was 20 minutes. The median fitting time for the LNR was 8 seconds (IQR $= 3$ seconds), and the total time for fitting the 40 datasets without parallelization was 6 minutes. The median fitting time for the full RDM was 16 seconds (IQR $= 7$ seconds), and the total time for fitting the 40 datasets without parallelization was 36 minutes.

**FINAL REMARKS**

In this article, we have introduced a computational framework for modeling predictive mechanisms in speeded decision making. Specifically, PAM combines computational accounts of predictive (Bayesian) processing with evidence accumulation models for decision making. Due to its efficiency and flexibility, it can represent a promising tool for the empirical investigation of how predictions and incoming sensory evidence interact and lead to decisions.

To validate our framework, we used parameter recovery via simulations. Specifically, we simulated a wide range of common scenarios for speeded decision-making tasks, with both fast and slow RTs, and high and low levels of accuracy. In all scenarios and with all the proposed decision models, the results showed highly accurate parameter recovery. Furthermore, PAM demonstrated good computational efficiency, which is crucial for practical applications. PAM is very fast when used with the powerful HGF toolbox, as well as with native optimization algorithms in MATLAB, Python, and R ([github.com/antovis86/PAM-PredictiveAccumulationModels](github.com/antovis86/PAM-PredictiveAccumulationModels)). In summary, PAM exhibited excellent performance in terms of both parameter estimation and computational time.

Another important feature of PAM is its flexibility, allowing researchers to tailor the framework to diverse experimental designs and research questions. Grounded in the "Observing the Observer" framework (Daunizeau, den Ouden, Pessiglione, Kiebel, Friston, et al., 2010; Daunizeau, den Ouden, Pessiglione, Kiebel, Stephan, et al., 2010) and the HGF toolbox (Frässle et al., 2021; Mathys et al., 2011, 2014), PAM supports the integration of various models in both perceptual and decision components, making it adaptable to different theoretical and empirical contexts. Here, we have demonstrated the use of PAM with two different perceptual models. Additionally, we have shown how different quantities from the perceptual model, specifically prior beliefs and their precision, can be used to modulate EAM parameters. Overall, the models presented here are just examples of how this framework can be used, chosen because they are suitable for common experimental designs. Nonetheless, PAM is not limited to these examples and is open to future developments and customizations. The GitHub repository already contains examples where other quantities, such as surprise, modulate EAM parameters, as well as an example for tasks requiring more than two possible responses. Moreover, we plan to implement a version suited for modeling perceptual decision-making tasks with different levels of sensory noise.

Delving into the theoretical implications, consider the two frameworks at the core of PAM, namely, the 'observing the observer' and EAM. To date, experimental research using these two computational approaches has proceeded largely independently. On one hand, modeling based on the 'observing the observer' framework has put significant effort into the development of perceptual model components that can capture various aspects of (perceptual) inference. At the same time, for decision model components, much simpler forms have been chosen (e.g., unit-square sigmoid, softmax function, simple linear regression-like models). This suggests that in this class of models, the emphasis is on describing inferential processes, while response models are used mainly to make inferences about perceptual models rather than to explain the dynamics of decision-making. Conversely, EAMs have been developed to accurately account for RT distributions of correct and incorrect decisions, providing meaningful psychological parameters that describe cognitive processes starting after stimulus encoding and continuing until response. Notably, more recently, these models have been applied to behavioral data from applied domains (as investigated in Human Factors research) to investigate the underlying cognitive processes (Boag et al., 2023). However, EAMs have largely overlooked the predictive nature of cognition. From these premises, our work should be understood as an effort to integrate these two powerful decision-making frameworks.

The relevance of a broader and more exhaustive investigation of the predicted mechanisms in decision-making emerges also from the results of the experiment described in the tutorial. Previous work suggests that prior beliefs about the target (expectations) can be incorporated into the Drift Diffusion Model (DDM) by adjusting the starting point, although some evidence indicates that expectations might modulate the drift rate (for an overview, see Summerfield & de Lange, 2014). Overall, our results are consistent with the second, less supported hypothesis, as the winning decision model was the RDM in which the drift rate $v$ was modulated by prior belief. Similarly, among the four DDM models the winning decision model was the one in which only $v$ was modulated by prior beliefs. It is clear that the results of a single experiment cannot provide a clear picture of the issue, especially considering the paucity of previous studies on which to base the interpretations. There are many aspects that need to be investigated and replicated. For example, in our RDK task, the accumulation was more decisional than sensorial since the stimulus was clear and presented for only 100 ms (Genest et al., 2016). But what happens when the stimulus duration is longer, allowing accumulating more sensory evidence? Or concerning the signal to noise ratio of the stimulus, how the precision of prior beliefs and sensory likelihood interact in the decisional process? These questions are just a hint of the lack of knowledge we need to fill.

In conclusion, the Predictive evidence Accumulation Models (PAM) offers a novel and efficient framework for investigating the interaction between predictive processing and sensory evidence in decision-making. Our simulations confirm PAM's accuracy and computational efficiency, making it a valuable tool for empirical research. This work contributes to advancing the field of cognitive science by providing deeper insights into the mechanisms underlying decision-making. We hope that PAM will foster new research and lead to a greater understanding of how predictions and sensory evidence interact in decision-making.

**Open Practices Statement**

The data and analysis scripts associated with this study are available on the Open Science Framework at https://osf.io/3jve9/. Additionally, the source code for implementing the Predictive Accumulation Models (PAM) framework, along with examples and tutorials, is publicly available on the GitHub repository at https://github.com/antovis86/PAM-PredictiveAccumulationModels. This study was not preregistered.

**Funding:** This work was supported by the STARS Grants program Starting grant (University of Padova, Italy) to A.V., and by the PRIN 2020 grant (protocol 2020529PCP) from the Italian Ministry of University and Research (MUR) to E.A. and F.D..